# The joint influence of topography and atmosphere on the meridional transport fluctuations in the Southern Ocean and its link with the ENSO events

Vladimir N. Stepanov

Proudman Oceanographic Laboratory, Merseyside, England

e-mail: vst@pol.ac.uk; tel.: 44(0)151 795 4859

February 19, 2007

## Abstract

This article deals with the investigation of the influence of topography and coastlines on the dynamics of the depth averaged Antarctic Circumpolar Current (ACC), driven by wind and atmospheric pressure. This is achieved with the help of a global barotropic circulation model, under idealized and real atmospheric conditions.

It is shown that the variability of meridional mass fluxes due to the atmospheric conditions over the ACC can induce short-period density anomalies in the Southern Ocean to the north in the vicinity of $47^oS$, which can be transferred to low latitudes by the wave mechanism described by Ivchenko et al. 2004, that could have significance with respect to rapid extra-tropical oceanic links. The results of the barotropic modeling demonstrate that changes in wind strength over the ACC, together with the effect of bottom topography and coastlines, induce some meridional flow variability in the Southern Ocean, and this meridional flux variability in the Pacific Ocean is anticorrelated with the strength of the wind over the ACC. The possible link between the short-term variations of the meridional flux in the Pacific sector and the ENSO events (El Niño Southern Oscillation) is discussed.

# 1. Introduction

Stepanov and Hughes 2006, hereinafter referred to as SH06, with the help of a barotropic model, studied the influence of topography and coastlines on the depth-averaged dynamics of the wind-driven Antarctic Circumpolar Current (ACC). They showed that there is a mass exchange between the Atlantic, Southern Ocean and Pacific basins both at short and long periods (~5 days and 30-100 days respectively). This exchange takes place with the global adjustment process at time scale about 30 days. The reason for the main exchange between the Southern Ocean and the Pacific is related to the balance of wind stress by form stress (a pressure difference across topographic obstacles) in Drake Passage. According to SH06, there are three major regions (the Drake Passage region, Kerguelan Plateau and the Pacific-Antarctic Rise) in the Southern Ocean that are responsible for approximately 65% of total form stress on the ACC; Drake Passage is the most significant topographic feature, accounting for about 30% of the total form stress. These three regions are shaded on figure 1. Eastward wind stress results in reduced bottom pressure adjacent to Antarctica, but at the same time, the form stress required to balance the wind stress results in greater bottom pressure on the western side of some topographic obstacle than on its eastern side. This together with the atmospheric forcing fluctuations over the ACC can result in to some meridional mass transport variability near bottom ridges and generate associated pressure or density (especially in the vicinity of a global front extending around the Southern Ocean) anomalies.

These signals due to topographic obstacles can be transmitted to more northern latitudes by the mechanism described by paper (Ivchenko et al. 2004, henceforth IZD04). They showed that signals due to salinity anomalies generated near to Antarctica can propagate almost without changes of disturbance amplitude in the form of fast-moving barotropic Rossby waves. Such waves propagate from Drake Passage (where the ACC is constricted to its narrowest meridional extent) to the western Pacific and are reflected at the western boundary of the Pacific before moving equatorwards and further northwards along the coastline as coastally trapped Kelvin waves. Such signals propagate from Drake Passage to the equator in only a few weeks and through the equatorial region in a few months. The authors suggested this mechanism could have significance with respect to rapid extra-tropical oceanic links. In a more realistic coupled ocean-atmosphere general circulation model, Richardson et al. 2005, observed a similar rapid response of the Pacific to a similar density anomaly in the ACC. The most important point of these results is the appearance of an initial barotropic disturbance in the Southern Ocean which propagates then into the tropical Pacific. The appearance of this initial barotropic signal is not dependent upon the source of the initial disturbance (since, e.g., either a change in wind stress or a density anomaly (see, e.g., Blaker et al. 2006) interacts with topography producing a barotropic signal).

SH06 developed further the idea from IZD04 about the influence of the Southern Ocean in the tropical Pacific and proposed that a wind-driven Southern Hemisphere Annular Mode (SAM) of variability (Hughes et al. 1999) could produce the barotropic signal, which propagates into the tropical Pacific, and subsequently interacts there with stratification (e.g., by means of internal (baroclinic) Kelvin waves) to produce temperature anomalies (the SAM consists of an oscillation of sea level atmospheric pressure between polar and subtropical latitudes, associated with oscillations in circumpolar winds).

The authors SH06 demonstrated also from idealized experiments that the Pacific response to Southern Ocean wind stress is close to the equilibrium response at periods longer than 30 days with a significant phase lag between them, although this response is more complicated at higher frequencies, suggestive of the excitation of free modes (that is strongly associated with closed $f/H$ contours around Antarctica, where $f$ is the Coriolis parameter and $H$ is the ocean depth) at periods shorter than about 10 days.

Thus, the mass exchange between the Southern Ocean and the Pacific at time scale of about 1-3 months, which is due to an association between the SAM and the form stress, can produce some short-period meridional flux variability in each sector of the Southern Ocean. This article considers the plausible distribution of this variability in the Southern Ocean due to the joint effect of topography, coastlines, wind and atmospheric pressure over the ACC. This variability leads to the appearance of some barotropic signal (due to either direct mass variability or via density anomalies) initiating barotropic wave propagation to the tropics. If barotropic waves are appeared then they can produce an amplified baroclinic response at the equator at least by means of two mechanisms. The first is a resonant interaction of the barotropic waves with baroclinic modes at the equator (Reznik and Zeitlin 2006). The second is that the barotropic adjustment process leads to currents along the continental slope which excite baroclinic modes by advection of density gradients; advection processes initiated by barotropic signal can be also substantial in the equatorial oceanic thermocline (that due to the propagation of internal Kelvin waves, see, e.g., Blaker et al. 2006). Therefore some correlation between short-period meridional flux variability in the Southern Ocean and the tropical variability should be observed in reality that will be considered by the paper too.

The meridional fluxes used in our analysis were obtained using a 1-degree resolution global barotropic model (Stepanov and Hughes 2004; Hughes and Stepanov 2004). These fluxes can result in the appearance of some short-period pressure or density anomalies in the Southern Ocean. How might the results of a barotropic model be analyzed to interpret the appearance of density anomalies in the real Southern Ocean? Figure 2 show "schematic profile" of the zonal average meridional component of the velocity at some latitude $\varphi$ in the Southern Ocean that could be obtained from the thermohaline circulation pattern. The meridional velocity is directed poleward in the upper thermocline, while lower water masses are moving equatorward. The value $V_o$ on the figure denotes the depth averaged meridional velocity, i.e., the barotropic constituent, which can vary depending on the atmospheric conditions. Willebrand et el. 1980, have shown from measurements that for time scales 10-300 days the ocean response to the applied external forcing (e.g., the wind stress) is barotropic, so the processes with time scale of a few months should be properly described by a barotropic model. The value $V_o$ corresponds to some meridional flow (denote it as $Q=\int (V_o (H+\zeta) d\lambda$, where $\zeta$ is sea level; $\lambda$ is longitude) describing the equilibrium state (say $Q \equiv Q_o$). If the meridional flux is increased, i.e., $Q=Q_o + Q'$ and $Q' > 0$, then more cold deep Antarctica Bottom Water (AABW) moves equatorward (fig. 2a) producing some cold water anomalies to the north in the vicinity of this φ and vice versa for the case $Q' < 0$ (fig. 2b), i.e., the barotropic model is able to provide the information about the appearance of some density anomalies in the Southern Ocean at the time scales of several months when the effect of barotropic changes has not yet transformed into baroclinic ones.

This conclusion is especially valid for the ACC since the baroclinic Rossby waves cannot propagate to the west in the ACC (far from Antarctica's shelf) because of its high latitude, relatively weak stratification, and strong eastward flow. A current having a velocity faster than the baroclinic Rossby wave speed approximates an equivalent-barotropic structure, in which the streamlines at depth are parallel to those at the surface (see, e.g., Killworth 1992; Killworth and Hughes 2002, etc.), i.e., the profile of current velocity is approximated by some self-similar solution depending on depth that leads to the monotonic dependence of 3-dimensional horizontal velocity on the depth averaged velocity.

Other confirmation of suitability to use barotropic results for the ACC was provided by Webb and de Cuevas. They investigated with the help of the OCCAM Global Ocean Model how the ACC responds to model changes in the wind stress at the latitudes of Drake Passage.

It was shown that the response is primarily barotropic which is controlled by topography, and only at the end of thirty days the baroclinic contribution to the momentum transfer reached 20% of the total (Webb and de Cuevas, submitted manuscript, 2006). The barotropic nature of future possible climatic changes in the Southern Ocean has been noted also in a global coupled climate modelling (see, e.g., Kim et al. 2005).

The aim of this paper is to investigate the variability of meridional fluxes in the Southern Ocean and to demonstrate the relationship between this variability and the ENSO events. To avoid some misunderstanding it should be mentioned that no processes connected with the changes of the meridional overturning circulation are considered in this paper because the use of barotropic model is able to provide only the information about short-time fluctuations of meridional mass fluxes at different latitudes due to the variability of atmospheric field. These fluctuations can lead to the appearance of some short-period density anomalies in the Southern Ocean in the vicinity of regions with high meridional flux variability at the time scales of several months, when the effect of barotropic changes has not yet transformed into baroclinic ones.

## 2. Method and Results

The transport processes in the Southern Ocean were studied with the help of a $1^o$ resolution global barotropic model (Stepanov and Hughes 2004). Main equations and parameters of the barotropic model are presented in Appendix. The model reproduces the large scale ocean variability quite well (Hughes and Stepanov 2004). The results of modelling for southern Atlantic (Hughes et al., submitted manuscript, 2006) showed that model results describing large scale ocean features are generally not dependent on a higher barotropic model resolution ($0.25^o$) or the choice of low bottom friction values, so 1-degree resolution is enough for the task considered here.

To check reliability of the barotropic model used, the ACC transport at Drake Passage calculated with the barotropic model was compared with the ACC transport obtained from the 1/4°–degree resolution OCCAM (3-dimensional) general circulation model (Webb and Cuevas 2002a) that was forced with the same ECMWF (the European Centre for Medium-Range Weather Forecasts) reanalysis meteorological data, but for the shorter duration of 8 years (see details in Hughes et al. 2003). The Drake Passage transport series are closely related to ACC transport at all longitudes in these two models, and also, the transport time series from these two models are highly correlated (correlation coefficient is 0.7 for daily values, significant at the 99% level).

Despite the lower resolution, and therefore different representation of topography in the barotropic model, the high correlation between the two model results confirms the conclusion that the ACC transport fluctuations are predominantly due to barotropic processes (see also Whitworth and Peterson 1985; Nowlin and Klinck 1986). Therefore we can use the barotropic model to investigate transport processes in the Southern Ocean at daily to intra-seasonal time scale (see also Meredith et al 2004). However, with the barotropic model we can only take into account variations due to wind and atmospheric pressure, disregarding thermohaline processes in the ocean, but it is consistent with the aim of the article.

The model was forced both with idealized model winds (henceforth, experiments A) and with 6-hourly winds and atmospheric pressures using ECMWF reanalysis from 1985 until the end of 2004 (experiment B). The use of the simplified conditions helps us to interpret the results more easily and can improve our understanding of the transport processes in the Southern Ocean, therefore we have used the idealized experiments that are forced only by zonal wind given by a zonal harmonic oscillation with the amplitude of 5 m/s (leading to a stress of 0.03 Pa) at all latitudes south of 40°S. These oscillations were independent of longitudes and latitudes and had the period of oscillations, $T_{os}$, varying from 2 to 35 days. The small value of wind stress (the mean is about 0.2 Pa) is supposed to represent a plausible amplitude for a fluctuation, rather than the mean.

The most dramatic sea level variations in both experiments are observed downstream of the shaded regions presented on fig. 1. Many of these "energetic" regions are coincident with closed *f/H* contours (see fig. 1) and are partly a result of topographic Rossby waves. As we can see from fig. 3, there are a few main regions of intense barotropic variability in the Southern Ocean found in all these experiments: southwest of Australia, west of South America, the Weddell Sea and the region to the south of South Africa. The existence of such "energetic" regions in the Southern Ocean (excluding the Weddell Sea) is confirmed by results obtained by other authors (e.g., Chao and Fu 1995; Fu and Davidson 1995; Fukumori et al. 1998; Mathers and Woodworth 2001; Webb and de Cuevas 2002 a, b), but their influence on the dynamics of the ACC has not been considered before. The locations of areas with the maximum sea level deflections in both experiments are mainly coincident with each other (figure 3a-c). There is elevated sea level far beyond the area where the idealized zonal wind stress is applied, especially a relatively large response (few millimeters) observed in the whole Pacific that is in detail discussed in SH06.

Apparent resonant behaviour can be seen from figure 3b-d in all sectors of the Southern Ocean at a period of more than about 5 days. These maps of sea level variation demonstrate an increase in its variability extending into the northern hemisphere from all sectors of the Southern Ocean for a forcing period from about 5 days and higher, i.e., one can see response of the almost whole World Ocean at these periods, while the exciting force is defined only in the southern hemisphere to the south of $40^oS$. The higher level of such variability persists in the Pacific Ocean for longer periods of the forcing (up to about 20 days), where the Pacific Ocean revealed a greater response than the other sectors that is consistent with the experiments using the real atmospheric forcings that were discussed by SH06.

The meridional fluxes from 60 to 30°S with a latitudinal step of 5° for each part Indian ($Q_I$), Pacific ($Q_P$) and Atlantic ($Q_A$) sectors of the Southern Ocean were calculated from all these model data (remember that there are no model external forces to the north of 40°S for the case of experiment A).

## 2.1 Model results for idealized forcings

The oscillations of meridional fluxes at all latitudes arise with dominant frequency corresponding to the period of $T_{os}$. More detail analysis shows that there are also weak oscillations with frequencies at three times the frequency of the external force due to the quadratic dependence of wind stress upon a wind speed that provides an excitation at this frequency. They have amplitudes about or less than 1/3 of those at period $T_{os}$ and the dependence of amplitudes and phase for these oscillations upon period $T_{os}$ is similar to the ones of main oscillations.

The oscillations of $Q_A$, $Q_I$, and $Q_P$ (in Sverdrup, 1 Sv=$10^6$ m$^3$/s) from experiment A for $T_{os}$=25 days depending on latitude and time are presented in figure 4 a-c. Diverging (converging) zones appear between 40 and 55°S in the Atlantic and Indian sectors of the Southern Ocean with a period determined by the forcing, i.e., north of this zone water masses are moving equatorward (poleward), while south of this poleward (equatorward) motion is observed. A similar zone of periodic divergence/convergence is observed at some periods in the Pacific sector too, but it is not as clear as in other sectors of the Southern Ocean. The removing of Drake Passage topography leads to blurring of such divergence/convergence zones in the Atlantic and Indian sectors, but these zones remain in the Pacific sector, although they are modified and displaced (fig. 4 d-f). It says that such behaviour in the Pacific is due to the interaction of wind with topography in the Pacific rather than in Drake Passage, while Drake Passage topography plays more important role in the Atlantic and Indian sectors.

The meridional fluxes in the western and eastern parts of each sector of the Southern Ocean were calculated too. These calculations demonstrate that the meridional fluxes in the western and eastern parts of the Pacific sector of the Southern Ocean are out of phase (this is valid for the Atlantic and Indian sectors too). The total and western mass fluxes have the same direction in the Indian sector for almost all frequencies of model atmospheric forcing, and in the Atlantic and Pacific sectors for $T_{os}$ less than 10-12 days, but the directions of the total and western meridional fluxes in the Atlantic and Pacific sectors for longer periods of the external forcings are often opposite. The fluxes in western and eastern portions of each sector are more than an order of magnitude greater than the total flux.

## 2.2 Model results for real forcings

A similar analysis of meridional flux variability has been made for the daily values of experiment B obtained from model runs with real atmospheric forcings. The annual cycle from all time series (7300 daily values) has been removed.

Experiments with only atmospheric pressure or wind stress have shown that variability due to both forcings are of the same order, so experiments driven only by a wind stress and by combined forcing (wind stress and atmospheric pressure) should be different; hence it is expected that the experiments with real forcings should be different from the idealized ones where only model wind fluctuations were used. The main discrepancy between experiments A and B is in the location of the mass diverging/converging zone detected in experiments with idealized forcing. In experiment B these zones are located between latitudes 47-48°S in all sectors of the Southern Ocean. The location of this zone coincides mainly with the boundary between regions where atmospheric cyclones to the south of 48°S and anticyclones to the north of 47°S move eastwards over the ACC, producing wind stress and atmospheric pressure fluctuations which have the effect of 'pumping' water masses for long distances to the north and south apart from this zone. The meridional fluxes for each sector in latitude bands 30-47°S and 48-60°S are highly correlated within each band but there is a significant anticorrelation between fluxes in the same sector of the Southern Ocean from the different latitude bands. However, there are some periods (usually the periods of boreal spring) when meridional flux fluctuations have the same directions in the whole latitude band, but the values of these fluctuations to the north of the latitude 47°S are smaller in comparison with ones to the south of 48°S (fig. 5 a presents the meridional flux in the Pacific sector depending on the latitude and time from 1985 to 2004; fig. 5 b, c present the same field but for two years in enlarged scale).

The second discrepancy is the directions of the total daily meridional flux and those from the western parts of each sector of the Southern Ocean. They have the same directions for all sectors of the Southern Ocean now, i.e., mass flows in western parts of each sector of the Southern Ocean are greater than for the eastern ones. The fluxes in western and eastern portions of each sector are more than an order of magnitude greater than the total (more than 20 Sv), and they could have a substantial influence on the appearance of density anomalies in the real Southern Ocean.

The detailed analysis for the exchange of atmospheric plus oceanic mass between ocean basins was presented by SH06; here the exchange of water between ocean basins will be discussed. The calculations of Indonesian throughflows and flow through Bering Straits demonstrate that they are more than an order and two orders of magnitude, respectively less than the mass fluxes from each sector of the Southern Ocean at latitude circle to the south of 30°S (or the difference between them), therefore these fluxes can be neglected in the basin mass exchange analysis and only a mass exchange through open boundaries in the Southern Ocean can be considered. Rather than focus on high frequency fluctuations of meridional flux in each basin at some latitude circle, the term

$M(t) = \rho_o \int^t Q_m(t)/\varphi \, dt$,

was chosen to investigate this water exchange, which describes the increment or loss of the basin total mass through their open boundaries at a moment $t$ (that is proportional to the basin sea level averages). The position of this open boundary (i.e., where the value of meridional flow $Q_m(t)/_\varphi$ is calculated) was taken at the latitude of 40°S, instead of 30°S which was considered in SH06 as a boundary between the Southern Ocean and other ocean basins. The reason is to be sure that the analysis results are due to the depth-averaged transport and not the Ekman transport, since the meridional fluxes at latitude 30°S are highly correlated with Ekman transports, while these correlations are not significant further south.

Cross-spectra (not shown) of the Atlantic $M_A(t)$ and Southern Ocean $M_{SO}(t)$ with the Pacific $M_P(t)$ demonstrate that the water mass exchange between these ocean basins is generally consistent with the total mass exchange described by SH06: the Atlantic and Pacific exchange at 4-6 day period (with significant exchange being seen also at longer periods); the Southern Ocean and Pacific exchange at periods longer than 30 days, becoming clearer still at periods longer than 100 days, with a phase difference near 180°.

The correlation coefficient between the monthly-averaged SAM-index (that is defined as the difference in the normalized monthly zonal-mean sea level pressure between 40°S and 70°S and characterizes the strength of the wind over the ACC; the SAM-index was generated from NOAA Climate Data Assimilation System reanalysis data) and $M_P(t)$, associated with the meridional flux $Q_P$ at 40°S, is -0.2, while correlation between $Q_P$ and the SAM-index is higher (-0.43). All correlation coefficients presented in the paper are statistically significant at the 99% level that was determined via an effective sample size, following Bretherton et al. (1999).

Thus, the fluctuation of the wind over the ACC causes the variability of the meridional flux in the Pacific part of the Southern Ocean to the north in the vicinity of 47°S, which is anticorrelated with the Atlantic sector of the Southern Ocean and the strength of the wind over the ACC. This is consistent with idealized experiments too, but there are of course, some discrepancies between them that are obviously due to the more complicated nature of the interaction of real atmospheric pressure and wind stress.

Thus, our experiments have shown that the strong (weak) wind over the ACC together with the effect of atmospheric pressure fluctuations provide a poleward (equatorward) meridional flux fluctuation in the Pacific sector of the Southern Ocean (that follows from the anticorrelation between the SAM-index and $M_P(t)$ or $Q_P$), which leads to the low (high) short-period transport fluctuation of the cold AABW to the north of 47°S, i.e., warm (cold) short-period anomalies of the deep waters can be produced north of this latitude (see the description for fig. 2 in the introduction).

## 3. Climate application of model results

It is well known that climate anomalies like ENSO, cause floods, droughts and the collapse of fisheries. The generally accepted view is that these effects are primarily the result of interaction between the ocean and the atmosphere in the tropics. It has been established that the onset of ENSO depends on equatorial wind anomalies in the western Pacific during the preceding spring and summer, though these wind anomalies can trigger the ENSO when the oceanic conditions in the tropical Pacific are favourable to the development of the ENSO (see, e.g., Lengaigne et al. 2004) and as Eisenman et al. 2005 demonstrated, these wind anomalies are partially stochastic and partially affected by the large-scale ENSO dynamics, rather than being completely external to ENSO.

Figure 6 presents a correlation coefficient (solid line) between the zonal wind stress over the Pacific averaged from June to September and winter's NINO4-index (see, http://climexp.knmi.nl) averaged during three months (from December to February). There are the high correlations both with the tropics and with the regions either over Trade Wind area or winds over the ACC, so it is unclear which is favoured candidate for the major role. For comparison, the dotted line on fig. 6 represents the correlation coefficient between the model summer's meridional flow in the Pacific Ocean averaged during three months (from July to September) and a winter's NINO4-index that is statistically significant for region from 45 to 35°S too.

There is now compelling evidence about possible coupled links between processes in northern and southern hemispheres with ENSO variability. There are many papers (e.g., Lau et al. 2005; Nicholls et al. 2005; Müller and Roecker 2006; Moknov and Smirnov 2006 etc.) demonstrating the influence of ENSO events on weather in northern hemisphere. However, Dong et al. 2006, have shown from global coupled ocean atmosphere modelling a reverse link between southern hemisphere and ENSO: the warm Atlantic phase of the Atlantic Multidecadal Oscillation is related to weaker ENSO variability, likely by means of "an atmospheric bridge that rapidly conveys the influence of the Atlantic Ocean to the tropical Pacific". For the southern hemisphere, models and observations demonstrate statistically significant correlations between the Antarctic and the tropics with leads and lags of order months. The analysis of observation data (Simmonds and Jacka 1995; Yuan and Martinson 2000; Kwok and Comiso 2002) demonstrated that Antarctic sea-ice distribution is strongly correlated with the ENSO variability on time scales of order a few months: there are correlations between ENSO and Antarctic sea-ice where ENSO either leads or lags Antarctic sea-ice variability. The last correlations Yuan and Martinson 2000, explained by some form of atmospheric teleconnection between Antarctica and the equatorial Pacific, but recent model studies (e.g., IZD04, Blaker et al. 2006) suggest that the ocean may be more important than previously considered in linking high latitudes and the equator on seasonal-interannual timescales. So, let us consider the idea that the wind driven processes in the Southern Ocean can substantially influence ENSO.

ENSO events are clearly identifiable in the tropical Pacific by June-July with the maximum phase of their development in the winter calendar months (the correlation coefficient between June-July and December-February averages of NINO4-index is 0.80). Figure 7 presents lead-lag correlation between $M_P(t)$ and NINO4-index as a function of calendar month. This figure shows that warming (cooling) events in the tropical Pacific are correlated with the equatorward (poleward) mass fluctuations in the Southern Ocean during almost whole year with lead about three months, however in March and especially in the winter-spring period of the southern hemisphere (from June to September) when the atmospheric field changes in the Southern Ocean are substantially large, the meridional mass fluctuations grow up and in turn, amplify the warming (cooling) in the tropics during next 4-6 months. Some imprint correlations between the equatorward (poleward) mass fluctuations in the Southern Ocean from June to October with ENSO events in May-December of a previous year are seen in fig. 7 also, that are due to long duration of these events (2-4 years).

The above described correlation can be explained by the coupled interaction of the tropics and the Southern Ocean. Upper ocean warming (cooling) in the tropics (that, e.g., could be associated with a seasonal cycle) leads to an enhanced (decreased) heating in the upper troposphere over the tropical ascending region in the Pacific. It means warmer (colder) air is transferred by the Hadley cell from the tropics to the descending regions in the subtropics that slows down (speeds up) here the atmospheric downwelling which then weakens (strengthens) wind over the Southern Ocean (the case of a low (high) SAM index). As we saw earlier in this paper, the weak (strong) wind over the Southern Ocean is associated with equatorward (poleward) mass flux in the Southern Ocean to the north in the vicinity of $47^oS$ that, as we see from figure 7, leads to the amplification of an ENSO signal.

The interaction between the tropics and the Southern Ocean depends on the stochastic processes of ocean-atmospheric interactions in these regions. A substantial role in these stochastic processes, as we see from figure 7, is due to the mass flux variability in the Southern Ocean associated with the changes in atmospheric forcings over the ACC, and would the interaction between the tropics and high latitudes lead to the ENSO event or usual seasonal variability, depends on the processes in the Southern Ocean.

The lag 4-6 months between the variability of $M_P(t)$ and NINO4-index in the period of a boreal winter-spring season seen in figure 7 agrees with the time estimate from IZD04 needed for temperature/salinity anomalies from the Southern Ocean to reach low latitudes by means of barotropic wave mechanism. As was mentioned above, the maximum development of warm or cold ENSO events usually occur in the winter calendar months, so relying on the time estimate of 4-6 months the mass variability $M_P(t)$ for the period from July to September need to be analysed.

Figure 8 shows the transport through Drake Passage and the variability of $M_P(t)$ due to meridional transport fluctuations through the latitude of 40°S in the Pacific Ocean, averaged for July-September from the model's 20-year time series. There is a strong coincidence between the minimums and maximums of $M_P(t)$ in the Pacific, and cold and warm ENSO events, respectively.

So, for the case of a low SAM index we have a high value of meridional flow fluctuations of the AABW from the Pacific sector of the Southern Ocean to the north of 47°S (the daily mass variability in the Pacific $M_P(t)/_{\varphi=40S}$ averaged for summer periods is more than 500 Gt (Gigatonnes), although the real water exchange between the Pacific and the Southern Ocean is more than an order of magnitude higher because of the link between the meridional total flux and mass flux in the western portion of the Pacific described earlier) and, hence, the anomalies of cold deep waters occur to the north of this latitude. The value of daily mass variability in the Pacific about 5000 Gt gives an estimate of the typical size of the signal arriving in the tropical Pacific. This signal is substantial, as a positive (negative) mass change corresponds to thermocline elevation (depression) in the tropics about 50 m over an area of 10 degree by 100 degrees in the period of 3 months. The signals from the Southern Ocean are transferred to low latitudes of the Pacific Ocean by the mechanism described in IZD04, where they subsequently interact with stratification that produce an elevation of the thermocline in the west equatorial Pacific that amplifies warm ENSO events; and vice versa: a low value of meridional flow fluctuations of the AABW from the Southern Ocean to low latitudes (i.e., the case for high values of the SAM index) causes the depression of the thermocline in the west equatorial Pacific and, hence, a cold ENSO event develops stronger. It is seen clearly from figure 8 that the El Niño events in 86-87, 91-92, 97, 2003-04 and La Niña in 88-89 and 98-2000, occurred when the peaks and minimums of the meridional transport fluctuations from the Southern Ocean in the preceding summer were observed (the meridional fluxes for the preceding years before the cold and warm ENSO events in 1988-89 and 1996-97 are presented correspondingly on figures 5b and 5c too). The correlation coefficient between the $M_P(t)/_{\varphi=40S}$ presented on fig. 8 and the winter's NINO4-index (dashed line on fig. 8) is 0.84. It should be mentioned that the correlation between the meridional flux $Q_P$ at this latitude averaged for the same summer periods and the winter's NINO4-index is also statistically significant (0.60). Thus, our results demonstrate that changes of the atmospheric conditions over the ACC might have the potential to influence climate over a much broader area and could be considered as the amplified mechanism of ENSO events. On the basis of dependence revealed between meridional transport fluctuations from the Southern Ocean and ENSO events, the ENSO was successfully modelled with the help of a simple model representing a classical damped oscillator, that confirms hypothesis about the importance of the coupled interaction of the tropics and the Southern Ocean (Stepanov, submitted manuscript, 2006).

It is interesting to note the coincidence between our results and those obtained by (Kim et al. 2005) on the basis of couple climate modelling. They showed that the changes of the ACC transport due to the weakness of zonal wind over the ACC lead to the cooling in the vicinity of the Pacific-Antarctic Rise that is occurring at all depths and have a barotropic nature. It is consistent with our result about the anticorrelation between the wind strength and the meridional flux fluctuations in the Pacific sector of the Southern Ocean.

The presented here results are in agreement with findings by Alvarez-Garcia et al. 2006 that identified three classes of ENSO events. The first two classes are characterized by well known paradigms: by the westward propagation of warm heat content anomalies north of the equator (so called the delayed oscillation mechanism) and local development of heat content anomalies in the northwest tropical Pacific, associated with overlying wind stress curl anomalies (the recharge-discharge oscillator model, though the recharge-discharge model includes the delayed oscillator as a particular case and both classes could be attributed to the subsurface memory paradigm). The third class characterizing by a relatively quick development of ENSO events (less than 9 months after the changes of the above mentioned plausible exciting forcings) was not be considered by authors because no explanations of some plausible mechanisms for such events were proposed before, and this paper is the first attempt to present the plausible mechanism for such rapid ENSO events.

## 4. Summary

The variability of meridional mass fluxes which depends on the wind and atmospheric pressure over the ACC has been studied. The results of barotropic modeling have demonstrated that changes in the atmospheric forcings over the ACC (for example, variability in the SAM), together with an effect of bottom topography via form stress, result in the appearance of the mass diverging/converging zone between latitudes 47-48°S and induce variations in meridional flows in the Southern Ocean. The meridional flux north of this zone, in the Pacific Ocean, is anticorrelated with the wind strength over the ACC. This variability in meridional mass fluxes can induce some additional short-period pressure/density anomalies in the Southern Ocean north of 47°S, which could be transferred to low latitudes by the mechanism described by IZD04. This leads to the possibility that the dynamics of the Southern Ocean can influence sea surface temperature in the tropical Pacific.

It was discussed a possible coupled link between the tropics and the Southern Ocean and how the wind driven processes in the Southern Ocean can substantially amplify ENSO variability. It seems very logical, but new research with the help of a complete baroclinic model of the ocean circulation is required to test this hypothesis of the influence of atmospheric conditions over the ACC on global climate at decadal scales.

**Acknowledgments**. This work was funded by the Natural Environment Research Council. Thanks to Chris Hughes for useful discussion, and to John Huthnance and Philip Woodworth for commenting on this manuscript.

**Appendix**

The continuity equation and two-dimensional momentum equations for barotropic variables (sea level, $\zeta$ and two depth averaged components of the velocity, $u$ and $v$) can be written as

$$\zeta_t + a^{-1}[m\,\partial_\lambda(Hu) + m\,\partial_\varphi(Hvm)] = 0, \tag{A1}$$

$$\partial_t u - fv = -m\,a^{-1}g\,\eta_\lambda - (H\,a)^{-1}[m\partial_\lambda(u^2 H) + \partial_\varphi(uv H)] +$$
$$+ H^{-1} A_M \Delta(Hu) + \tau_\lambda/(H\rho_o) - \varepsilon_o H^{-1}|V|u, \tag{A2}$$

$$\partial_t v + fu = -a^{-1}g\,\eta_\varphi - (H\,a)^{-1}[m\partial_\lambda(uvH) + \partial_\varphi(v^2 H)] +$$
$$+ H^{-1} A_M \Delta(Hv) + \tau_\varphi/(H\rho_o) - \varepsilon_o H^{-1}|V|v, \tag{A3}$$

$$\eta = p_a/(\rho_o g) + \zeta(\lambda, \varphi, t), \tag{A4}$$

where $\varphi$ is latitude; $\lambda$ is longitude; $t$ is time; $g$ is gravity; $\boldsymbol{\tau} = (\tau_\lambda, \tau_\varphi)$ is a vector of wind stress; $p_a$ is the atmospheric pressure; $\rho_o$ is a reference density for water; $a$ is the radius of the Earth;

$m = \sec\varphi$; $f = 2\Omega \sin\varphi$; $\Omega$ is the Earth's rotation rate; $\Delta$ is Laplace operator; $H$ is ocean depth; $\varepsilon_o = \chi\varepsilon$, $\chi = 1 + (A_M/k_M/H)/|\nabla H|^2$; $k_M$ and $A_M$ are correspondingly vertical and lateral turbulent viscosity; $|V| = (u^2 + v^2)^{1/2}$ is a absolute value of the flow velocity.

A quadratic bottom friction law with coefficient of proportionality of ε approximates the bottom friction in the model. Finally, the value of the bottom friction in the model is modulated by gradient of the bottom topography ∇H through the dimensionless coefficient χ. This coefficient results from the integration of the horizontal viscosity term of the initial three dimension momentum equations and effect described by this term can be substantial on a coarse model grid.

The system of the momentum equations of the barotropic model should be completed by the lateral boundary condition: $u = v = 0$.

The following parameters of model were used in model runs: $\varepsilon = 0.03$; $A_M = 5 \times 10^3$ m$^2$/s; $(A_M/k_M/H) = 5 \times 10^5$; the timestep $\Delta t = 60$s.

# List of Figure Captions

**Fig. 1** The map of *f/H* contours and three major regions in the Southern Ocean (shaded areas) at the Drake Passage latitudes that are responsible for approximately 65% of total form stress on the ACC.

**Fig. 2** The schematic profile of the zonal averaged meridional component of the velocity in the Southern Ocean; a) is a case of high meridional flow and b) – when the flow is low. Solid arrows correspond to disturbed profiles of velocity and thin ones correspond to an equilibrium state.

**Fig. 3** The standard deviations of sea level for real atmospheric forcings (a), and for harmonic excitation at periods $T_{os}$ = 15, 7 and 4 days (b, c, d), respectively.

**Fig. 4** The fluctuations of the meridional fluxes in the Atlantic (a), Indian (b) and Pacific (c) sectors of the Southern Ocean for the case of $T_{os}$=25 days, depending on the latitude and model time from 20 to 60 days in the experiments A; d, e, f – are the same for the experiments with flat topography in Drake Passage.

**Fig. 5** The fluctuations of the meridional flow in the Pacific sector of the Southern Ocean depending on the latitude and time from 1985 to 2004 (a); b and c - are the same for the preceding years before the cold (1989) and warm (1997) ENSO events, respectively.

**Fig. 6** The correlation coefficient: solid line for the zonal wind stress over the Pacific sector of the Southern Ocean averaged from June until September and dotted line for the model summer's meridional flow in the Pacific Ocean averaged from July until September, each with the winter's NINO4-index (averaged during three months from December until February). Dashed line corresponds to the scaled profile of zonal wind stress that was averaged on time and zonally.

**Fig. 7** Lead-lag correlation between $M_P(t)$ and NINO4-index (3-months running averaged data for 1985-2004 were used) as a function of calendar month. Positive (negative) lag refers to case when $M_P(t)$ leads (lags) the NINO4-index. The values of correlations higher than 0.46 are statistically significant at 95% level.

**Fig. 8** The values of transport through Drake Passage in Sv (thin solid line) and variability of $M_P(t)$ due to meridional transport fluctuations through the latitude of 40°S in the Pacific Ocean in Gt (Gigatonns) (thick solid line) averaged for July-September. Symbols EL and LA denote warm and cold ENSO events, respectively. Dashed line corresponds to scaled winter's NINO4-index.

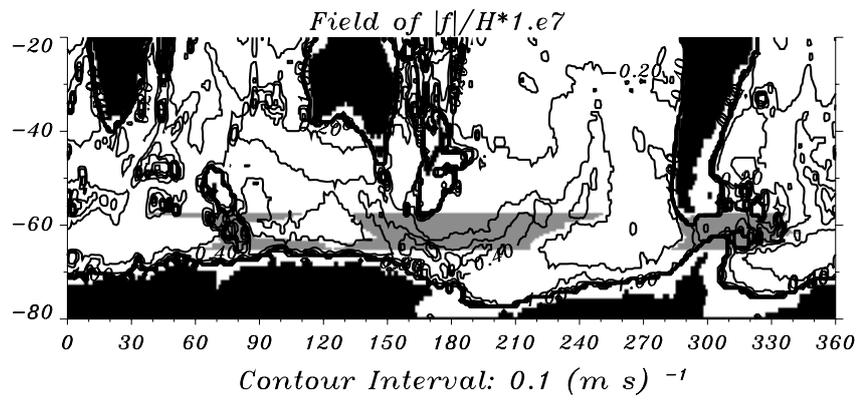

Figure 1

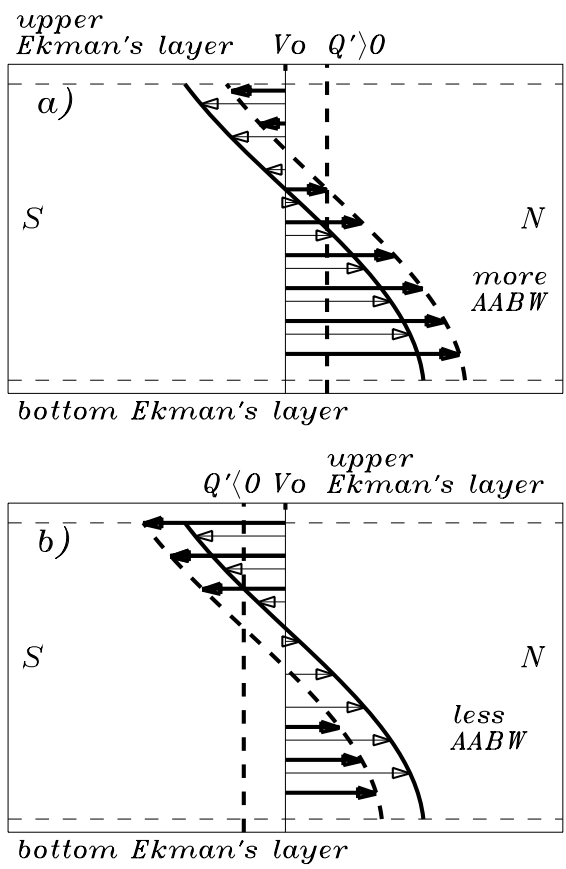

Figure 2

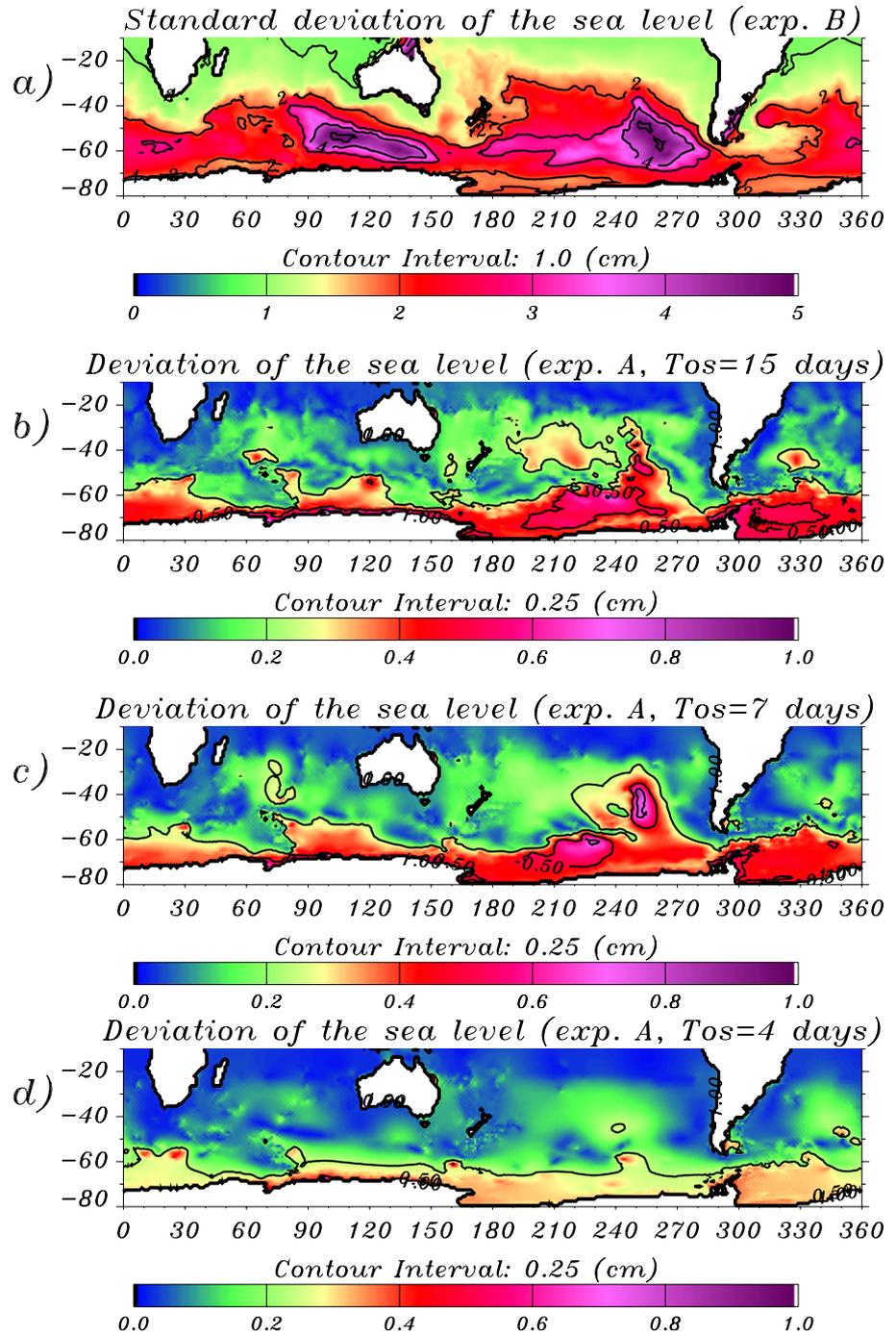

Figure 3

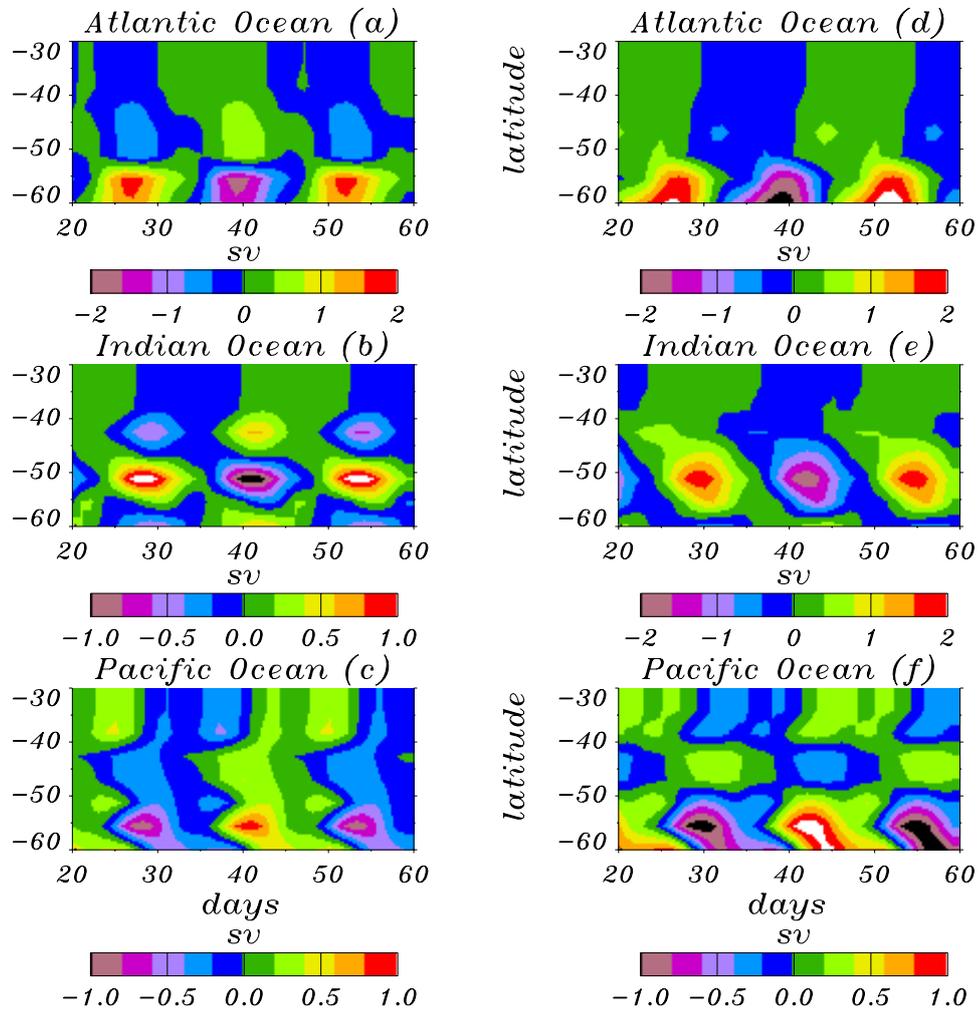

Figure 4

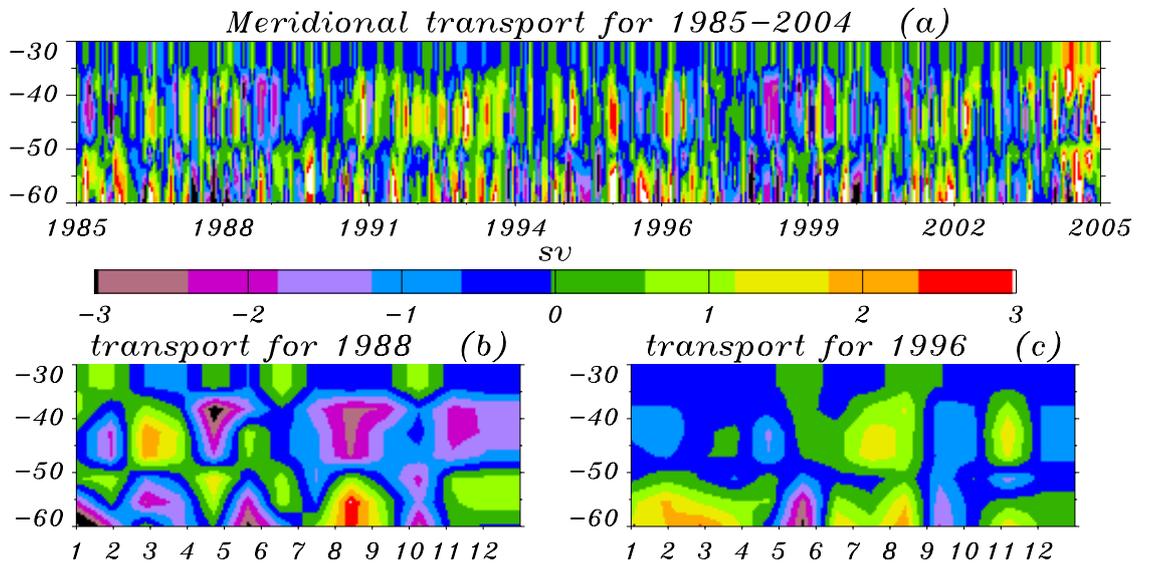

Figure 5

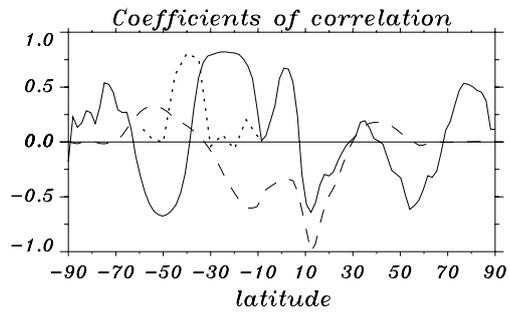

Figure 6

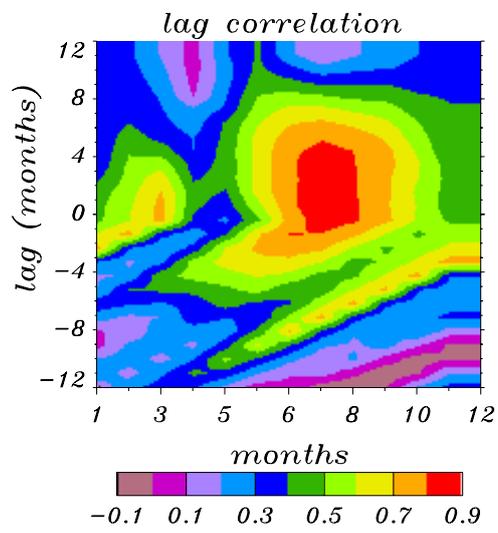

Figure 7

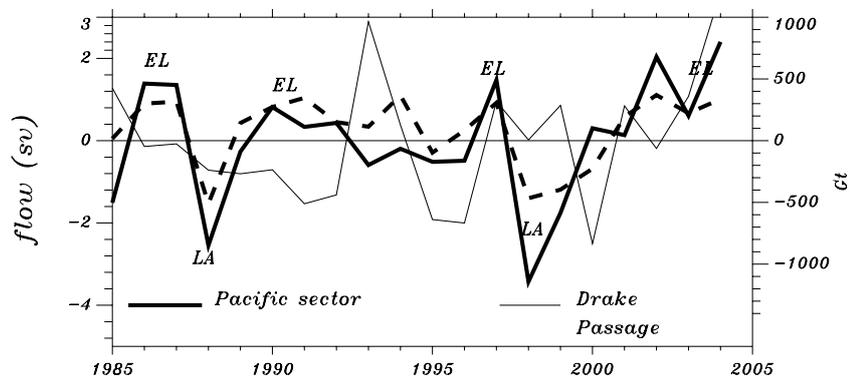

Figure 8